\documentclass[epj]{svjour}

\usepackage{graphicx}
\usepackage{pslatex}
\usepackage{txfonts}
\usepackage[T1]{fontenc}
\usepackage[latin1]{inputenc}
\usepackage{float}

\begin{document}
%
\title{Exploring the LHC medium with electromagnetic probes}
\author{A. Mar\'{\i}n \inst{1} for the ALICE Collaboration 
}                     


\institute{GSI, Planckstrasse 1, 64291 Darmstadt, Germany}

\date{Received: date / Revised version: date \today}

\abstract{
Heavy-ion collisions will enter a new era with the start of the CERN Large Hadron Collider
(LHC). A first short run with proton-proton collisions at the injection energy of 0.9 TeV will be 
followed by a longer one with $pp$ collisions at 10 TeV. First Pb-Pb collisions at $\sqrt{s_{NN}}$=5.5 TeV will take 
place in 2009. Three experiments (ALICE, ATLAS, and CMS) will study both $pp$ and Pb-Pb collisions.
A selection of results showing the capabilities of the three experiments for the study 
of the LHC medium with electromagnetic probes is presented. 
} 
\maketitle
\section{Introduction}
\label{intro}
With the startup of the CERN LHC heavy-ion collisions will enter a new era. With a center of mass
energy of 5.5 TeV, about a factor 30 larger than the available energy at RHIC, matter will be produced
in a new domain of high energy density (15 to 60 GeV/fm$^3$) and high initial 
temperature (factor 3 to 5 higher than the critical Temperature, T$_c$). At high temperature 
and high energy density QCD calculations \cite{karsch1,lat2} 
predict a phase transition from hadron gas to quark gluon plasma (QGP) and chiral symmetry restoration.
 The expected time the system will expend in the QGP phase is of the order of 10~$fm/c$ and it is expected to extend 
over a large volume. The cross sections of hard probes are increasing by large factors at LHC energies compared
to RHIC or SPS \cite{fabjan1}.
Among the different observables photons and dileptons are particularly interesting because they would directly probe 
the high temperature and high density phase of these collisions.
The pp runs will provide a test of pQCD as well as important reference data for heavy-ion runs. With the first 
heavy-ion run expected in 2009, the global event characteristics and bulk properties will be established, 
and hard probes measurements could be started.

\section{Direct photon phenomenology at the LHC}
There are different production mechanisms of photons in heavy-ion collisions.
Direct photons are produced at leading order (LO) in quark gluon  Compton scattering (qg$\rightarrow$ q$\gamma$)
and quark anti-quark annihilation (q$\bar q$ $\rightarrow$g$\gamma$). The next-to-leading order (NLO) process
is dominated by bremsstrahlung and fragmentation photons (qg$\rightarrow$qg$\gamma$). In addition, there are 
photons from the jet re-interaction in the medium and thermal photons from QGP and from the hot ha\-dro\-nic 
stage following the QGP. 
Direct photons have to be isolated from the main source of background photons, i.e. the decay of hadrons 
(mainly $\pi^0$'s and $\eta$'s) after the QGP phase.

\begin{figure}
\includegraphics[width=0.99\linewidth]{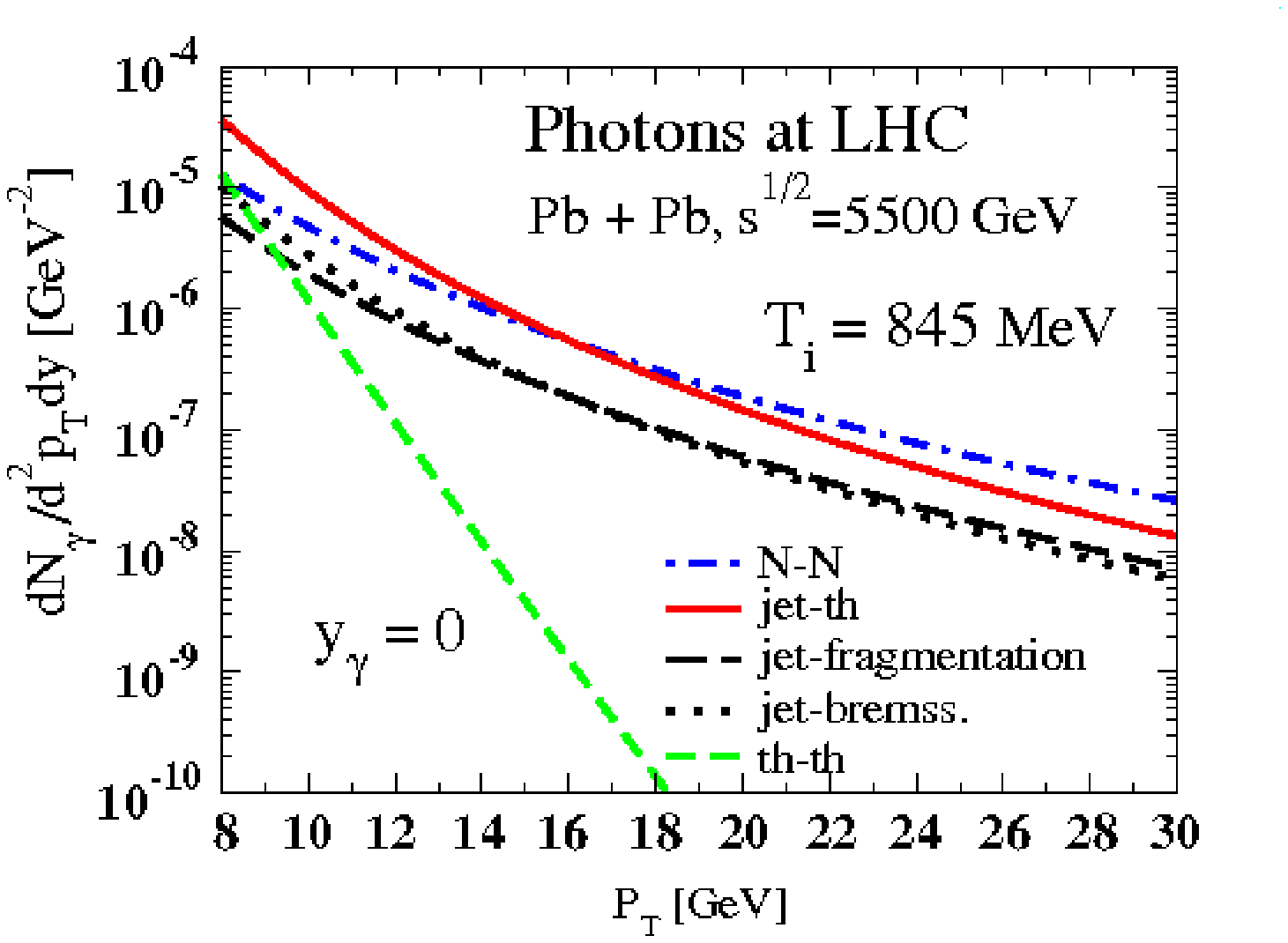}
\caption{Contributing sources of high-p$_T$ photons at midrapidity in central Pb-Pb collisions at LHC from \cite{phot1}.}
\label{figgamlhc}
\end{figure}

The different contributions to the direct photon spectrum for LHC energies as calculated by \cite{phot1} 
is presented in Fig.~\ref{figgamlhc}. Other predictions are also available \cite{phot2,phot3,phot4,armesto}. 
At RHIC $\gamma$'s from the parton scattering at LO were the dominant contribution 
for p$_T> 5$ GeV/c while at LHC jet-photon conversion in the plasma dominates between 8 and 14 GeV/c. 
Only for p$_T>$20 GeV/c the hard NN scattering dominates.
Although photons are more abundantly produced at LHC compared to RHIC (about a factor 10 larger), 
the ratio of direct photons to $\pi^0$ is smaller at LHC. The ratio is $\sim$10\% for a p$_T$ of 20 GeV/c \cite{photyellow}, 
thus a very good PID to distinguish between direct and decay photons would be essential. 

Recently it has been proposed \cite{turbide1} that in order to distinguish between the different 
photon sources one can study their azimuthal anisotropy as function of p$_T$ in conjuction with R$_{AA}$. 
A calculation for semicentral Pb-Pb collisions at LHC energies \cite{liu1} predicts
v$_2$ = 0 for initial production, positive for jet fragmentation and negative for jet photon conversion proceses. 
Therefore, a negative v$_2$ together with R$_{AA}>$1 will be an unambigous signature of medium produced photons.

The high p$_T$ photon measurement opens the possibility of studying the jet fragmentation function (FF) and its 
modification (redistribution of the parton energy inside the hadron jet) due to the medium effects in heavy-ion collisions
by using the $\gamma$-jet channel. As the photon is not affected by final state interactions, its transverse energy (E$_T$)
gives the energy of the jet before modification in the medium. This measurement would be complementary to jet 
reconstruction in the energy range from 20 GeV to 80 GeV \cite{klein1}, or above 75 GeV \cite{loizides1,grau1}.

\section{ALICE, ATLAS and CMS as photon and dilepton detectors at the LHC}

\begin{figure}
\includegraphics[width=0.95\linewidth]{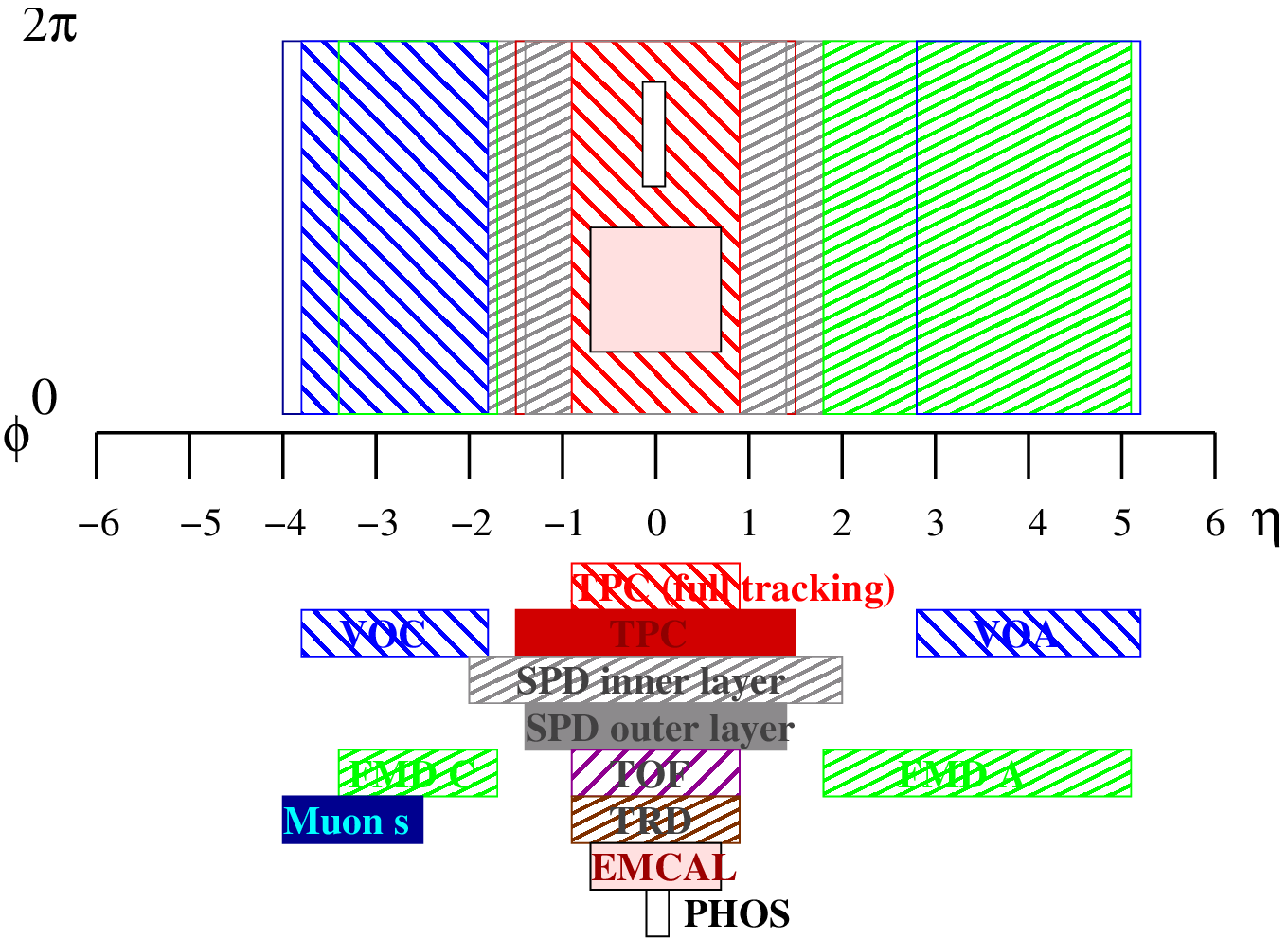}
\includegraphics[width=0.95\linewidth]{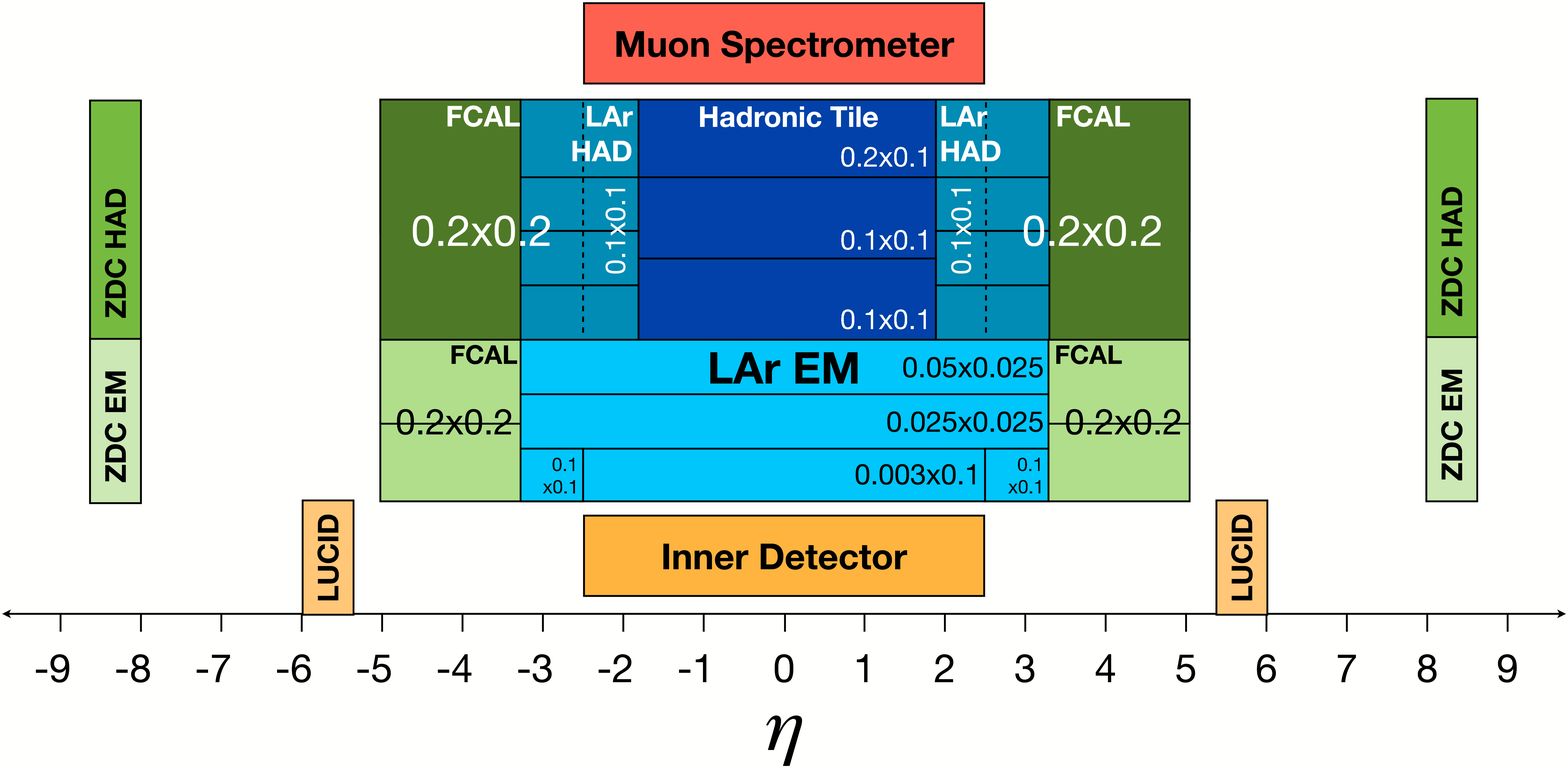}
\includegraphics[width=0.95\linewidth]{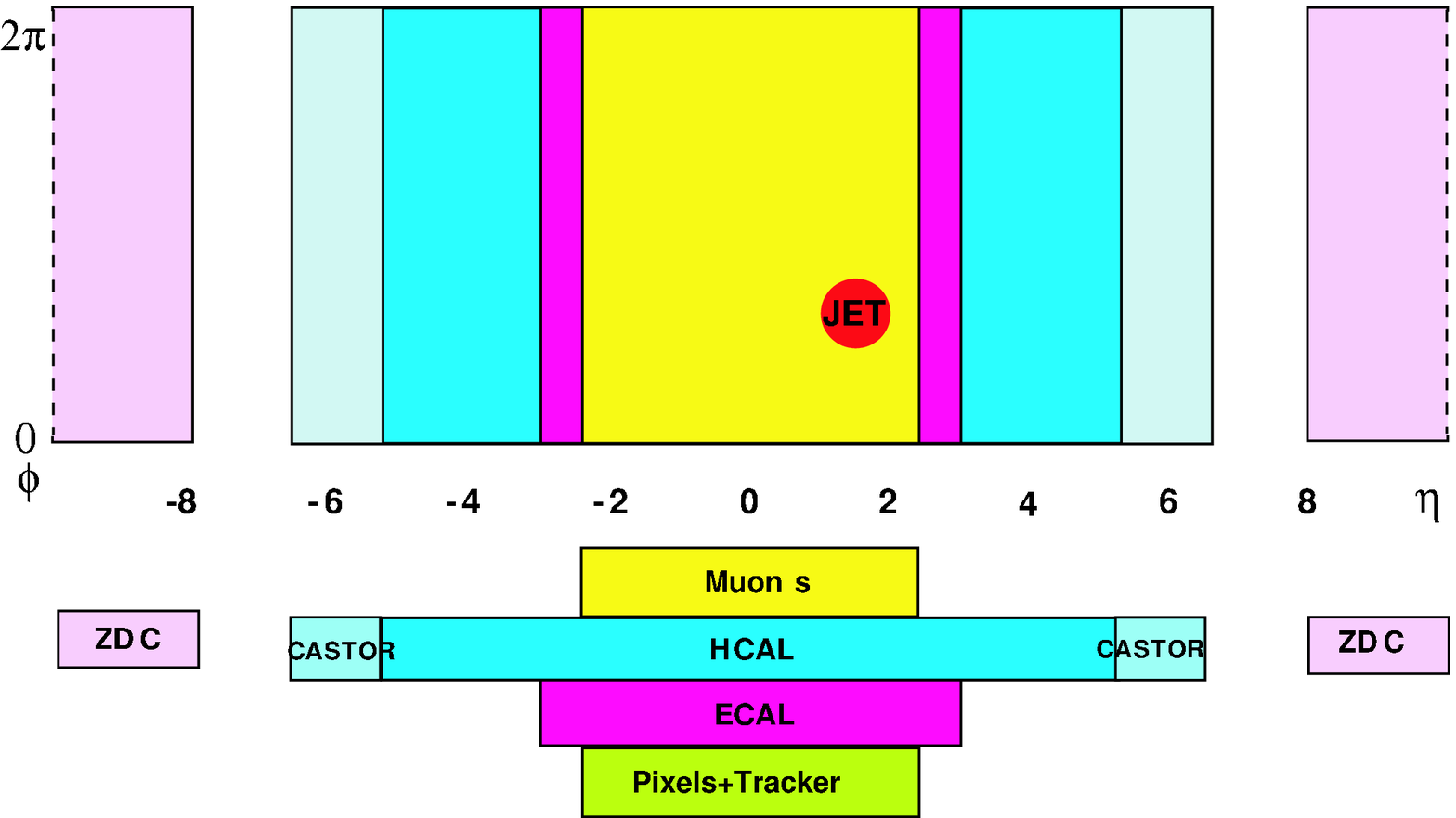}
\caption{ALICE (top), ATLAS (middle), and CMS (bottom) pseudorapidity and azimuthal coverage.}
\label{figpseudet}
\end{figure}

Three experiments will participate in the heavy-ion program at the CERN LHC.
ALICE is the experiment devoted to the study of heavy-ion collisions \cite{ali1,ali2}, 
and it will also participate in the pp running with an extensive physics program.
ATLAS \cite{atlas1} and CMS \cite{cms1} are experiments dedicated to study $pp$ collisions although by now 
they also have a rich heavy-ion program \cite{atlas2,cms2}.

\begin{figure*}
\includegraphics[width=0.3\linewidth]{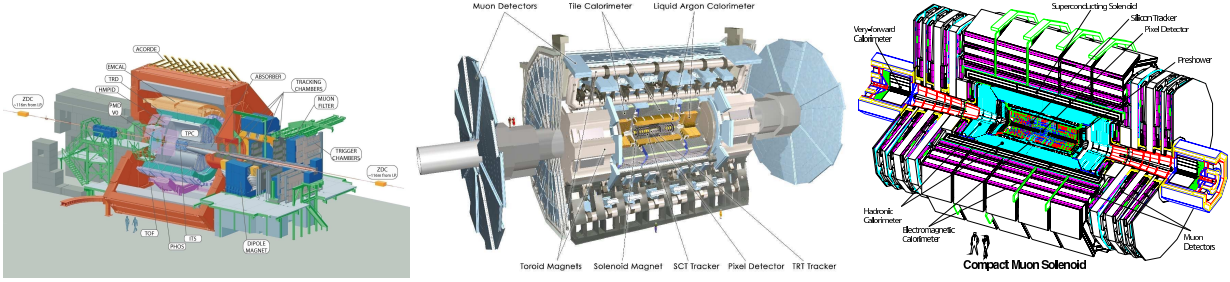}
\includegraphics[width=0.6\linewidth]{cms_atlas2.epsi}
\caption{Schematic layout of the ALICE (left), ATLAS (middle) and CMS (right) detectors.}
\label{figdet}
\end{figure*}

\begin{table*}
\begin{tabular}{|c||c||c||c||} \hline
~~~~~~Exp~~~~~~~&~~~~~~~~~~~~~~~~~ATLAS~~~~~~~~~~~~~~~~~~&~~~~~~~~~~~~~~~~~~~~~CMS~~~~~~~~~~~~~~~~~~~&~~~~~~~~~~~~~~~~~~~~~~~~~~~~~~~ALICE~~~~~~~~~~~~~~~~~~~~~~~~~~~~~~~~~~~~\\ \hline
\end{tabular} 
\begin{tabular}{|c||c|c||c|c||c|c|c||} \hline
Name     & LAr Barrel & LArEndCap & ECAL(EB) & ECAL(EE) & PHOS & EMCAL & Barrel \\ \hline
Coverage & 0$<|\eta|<$1.4& 1.4$<|\eta|<$3.2&  0$<|\eta|<$1.5& 1.5$<|\eta|<$3.&0$<|\eta|<$0.12& 0$<|\eta|<$0.7 & (0$<|\eta|<$0.9 \\ 
         &    2$\pi$     &     2$\pi$      &        2$\pi$  &     2$\pi$     &   0.6$\pi$    &  0.6$\pi$      &  2.$\pi$) $\cdot$ 7X/X$_0$ \\ \hline  
Granularity&0.003$\times$0.100&0.025$\times$0.100&0.017$\times$0.017&0.017$\times$0.017&0.004$\times$0.004&0.014$\times$0.014&3$\cdot$10$^{-4} \times$ 2$\cdot$10$^{-4}$\\ 
$\Delta\eta\times\Delta\phi$& 0.025$\times$0.025& 0.025$\times$0.025&  &  to &  & & resolution \\ 
           & 0.025$\times$0.025&0.025$\times$0.025& & 0.05 $\times$ 0.05 &  &    &   \\ \hline
Resolution &10\%/$\sqrt{E}$$\bigoplus$&10\%/$\sqrt{E}\bigoplus$&2.7\%/$\sqrt{E}\bigoplus$&5.7\%/$\sqrt{E}\bigoplus$&1.3\%/$\sqrt{E}\bigoplus$&7\%/$\sqrt{E}\bigoplus$&2\% low pt \\ 
           & 0.5\%  & 0.5\% & 0.55\% &0.55\% & 1.1\% & 1.5\% & 5\% high pt \\ \hline

\end{tabular}
\caption{Compilation of the photon detectors in the three  LHC experiments with a brief description of their characteristics.}
\label{tablephotons}
\end{table*}

The ALICE detector (Fig.~\ref{figpseudet} top , Fig.~\ref{figdet} left) \cite{ali1,ali2} consist 
of central barrel detectors, a forward muon spectrometer and forward multiplicity and centrality detectors. 
 The central barrel inside a large solenoid magnet with the field up to 0.5 T  covers almost 2 units 
of pseudo-rapidity ($|\eta|<$0.9). The charged particle tracking detectors are the ITS (Inner 
Tracking System) with three different silicon technologies (pixels, drift and strips), the 
TPC (Time Projection Chamber) main tracking system for charged particles, and the TRD (Transition 
Radiation Detector). All three tracking detectors are also used for particle identification (PID) by 
measuring ionization losses or transition radiation. The TOF (Time Of Flight) detector is used for 
PID in the intermediate momentum range (0.2 to 2.5 GeV/c). 
Three smaller single arm detectors complete the central barrel, the HMPID (High Momentum Particle 
Identification) detector that will extend the PID up to momenta of 4-5 GeV/c, the high resolution 
high granularity electromagnetic calorimeter PHOS (PHOton Spectrometer) to measure photons and 
neutral mesons and the EMCAL (ElectroMagnetic CALorimeter) to improve the ALICE capabilities in 
the measurement of high energy jets and direct photons.  
At forward rapidities, ALICE triggers and detects muons using the muon spectrometer with tracking and trigger 
chambers in a 3~Tm magnetic field. 
ALICE has an excellent PID: $\pi$/K/p identification up to 50 GeV/c and $\gamma$,$e$, and $\mu$ up to 100 GeV/c.
The ALICE specific photon detectors are PHOS and the EMCAL. In addition, photons that 
convert in the detector material ($\gamma$Z$\rightarrow$e$^-$e$^+$Z ) can be detected by measuring 
the e$^+$ and e$^-$ in the central barrel. This is a clean photon identification method, providing 
directional information used to reject non vertex background. The momentum resolution at low p$_T$ 
using this method is better than electromagnetic calorimeters. 
The possibility of using a L1 TRD trigger \cite{trd} to make use of this method at high p$_T$ is under investigation.

The ATLAS detector (Fig.~\ref{figpseudet} middle, Fig.~\ref{figdet} middle) \cite{atlas1,atlas2} has 
hermetic azimuthal coverage over a wide range in pseudo-rapidity. The unique feature of the ATLAS detector 
is its ca\-lo\-ri\-me\-try, composed of several independent longitudinal sampling layers of electromagnetic and 
hadronic calorimetry with full azimuthal and $|\eta|<$5 units coverage. Specially the Liquid Argon (LAr) 
electromagnetic calorimeter, which provides
excellent energy and position information on electrons and photons. The inner tracking system 
within a 2T solenoidal field is equiped with silicon pixels, silicon strips and straw-tube tran\-si\-tion-ra\-dia\-tion 
tracker for reconstructing charged tracks within  $|\eta|<$2.5. A muon spectrometer ($|\eta|<$3) is 
located within a toroidal field outside of the hadronic calorimeters.

The CMS detector (Fig.~\ref{figpseudet} bottom, Fig.~\ref{figdet} right) \cite{cms1,cms2} is also a hermetic detector with a 
large acceptance both for tracker and calorimetry. 
The CMS detector is a 22 m (length) x 15 m (diameter) detector featuring a 4 T solenoid surrounding central
silicon pixel and micro-strip tracking detectors ($|\eta|<$2.4) and electromagnetic ($|\eta|<$3) and hadronic 
($|\eta|<$5) calorimeters. Muon detectors ($|\eta|<$2.4) are embedded in the flux return iron yoke of the magnet. 
CMS detects leptons and both charged and neutral hadrons.
  
A compilation of the different photon detectors with their main characteristics in the three LHC experiments 
is presented in Table~\ref{tablephotons}. PHOS and LAr (Barrel) have the best granularity.

\section{Prompt photons and photon-tagged jets}
To demonstrate the direct photon reconstruction capabilities ALICE has generated events using PYTHIA \cite{pythia}
triggered by prompt photons ($\gamma$-jet events) or $\pi^0$ (jet-jet events). To simulate Pb-Pb events, the pp events
were merged with heavy-ion events generated by HIJING \cite{hijing}.
Prompt photons are selected in PHOS by applying cuts on the shape of the shower developed on the detector and
isolation cuts \cite{nimali1,nimali2}.
 
\begin{figure}[hbt]
\includegraphics[width=0.95\linewidth]{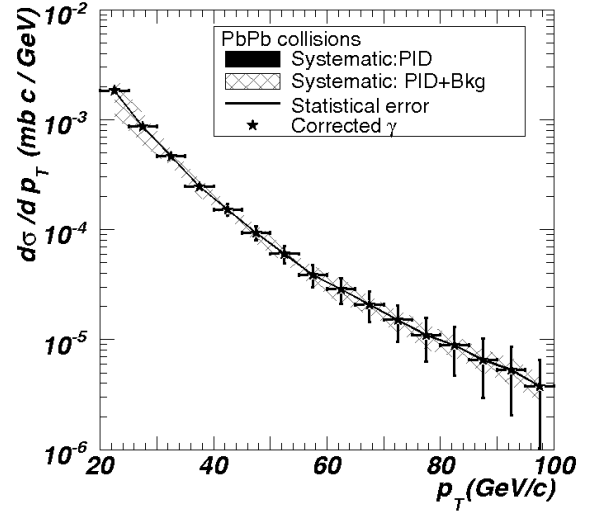}
\caption{Simulated prompt photon spectrum expected to be measured in ALICE during a LHC running year for 
Pb-Pb collisions with statistical (bars) and systematic (shaded band) errors.}
\label{figaligamPbPb}       
\end{figure}

The spectrum of direct photons detected in PHOS for Pb-Pb collisions \cite{nimali1,nimali2} is shown 
in Fig.~\ref{figaligamPbPb}. An isolation cut of radius R=$\sqrt{\Delta \eta^2 + \Delta\phi^2}$=0.2 and 
a p$_T$ threshold of 2 GeV/c is used. The background rejection with the applied cuts is 1/14 for an efficiency 
of 50\%. A sample of 2000 $\gamma$ with p$_T$>20 GeV/c is expected during one LHC running year. 

\begin{figure}[hbt]
\includegraphics[width=0.5\textwidth]{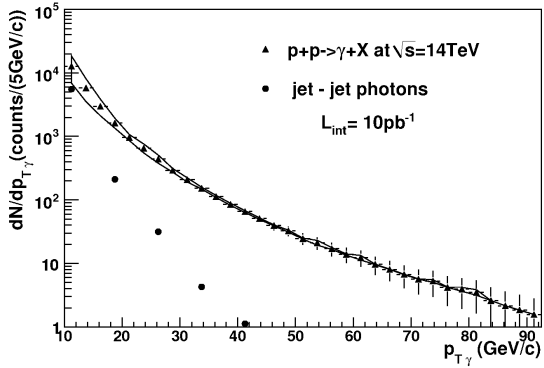}
\caption{Isolated photon spectrum that can be measured in ALICE with 2 PHOS modules in pp collisions at 14~TeV. 
The error bars are statistical errors. Systematic errors are given by the area around the data points.}
\label{figaligampp}       
\end{figure}

The spectrum of prompt photons detected in 2 PHOS modules for pp collisions is presented in Fig.~\ref{figaligampp}.
An isolation cut with R=0.3 and $\sum p_T <$~2~GeV/c has been used. A background rejection of 1/170 is obtained 
for an efficiency of 69\%. A sample of 3000 $\gamma$ with p$_T>$ 20 GeV/c is expected during one LHC 
running year (10~pb$^{-1}$) \cite{yaxian1}.

\begin{figure}[h]
\begin{center}
\includegraphics[width=0.5\textwidth]{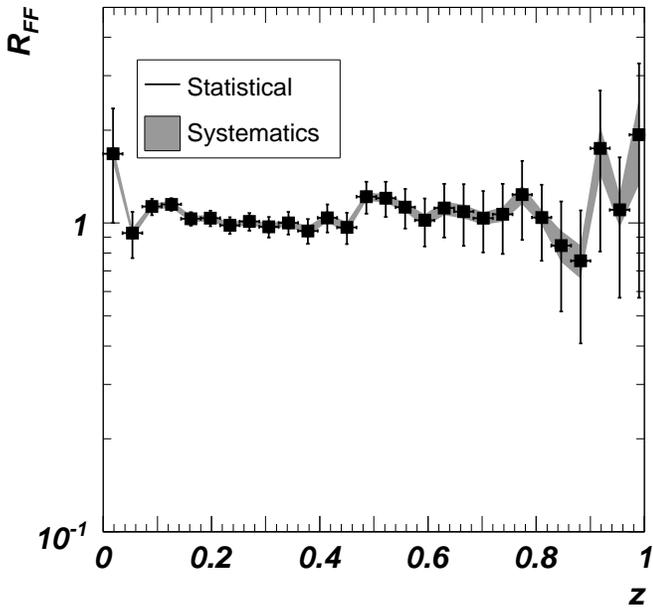}
\end{center}
\caption{Ratio of the fragmentation functions of $\gamma$-tagged jets with energy larger than 20 GeV for 
Pb-Pb collisions scaled to pp collisions detected in the central tracking system and EMCal. The shaded 
region represents the systematic error due to the contamination from jet-jet events.}
\label{figalicerffPbPb} 
\end{figure}

As direct photons are not perturbed by the medium the fragmentation function of the recoiled jet can be measured
using E$_\gamma$ as the jet energy. The ratio of the measured fragmentation functions for the two collision systems
shows that for E~$>$~20~GeV variations of R$_{FF}$ larger than 5\% in the range 0.1$<z=p_T/E_{jet}<0.5$ can be 
measured, see Fig.~\ref{figalicerffPbPb}.


The expected p$_T$ reach of the prompt photon spectrum will be larger in ATLAS and in 
CMS compared to ALICE due to their larger acceptance.
\begin{figure}[hbt]
\includegraphics[width=0.95\linewidth]{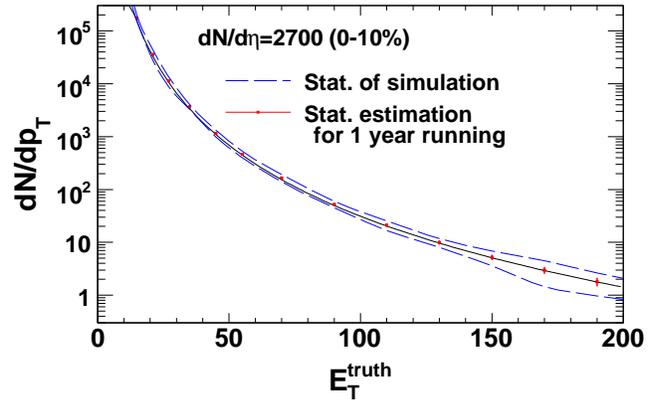}
\caption{Isolated photon spectrum as expected to be measured by the ATLAS detector in one LHC year running \cite{cole1}.}
\label{figatlasgamet}  
\end{figure}

Due to the fact that the ATLAS detector has very good granularity in the first layer, isolated photons give 
a very clear signal even in a heavy-ion background simulated using HIJING. The set of cuts based on shape of the signal
and isolation cuts used for this study give an efficiency of about $\sim$60\% and a rejection factor 
of $\sim$10 assuming R$_{AA}^h$/R$_{AA}^\gamma$ =1.
The direct photon spectrum resulting from combining shape and 
isolation cuts is presented in Fig.~\ref{figatlasgamet}~\cite{grau1,cole1}. 
A sample of about 100K photons with E$>$30~GeV and of about 10K with E$>$70~GeV are expected
for an integrated luminosity of 0.5 nb$^{-1}$.  

\begin{figure}[htb]
\includegraphics[width=0.95\linewidth]{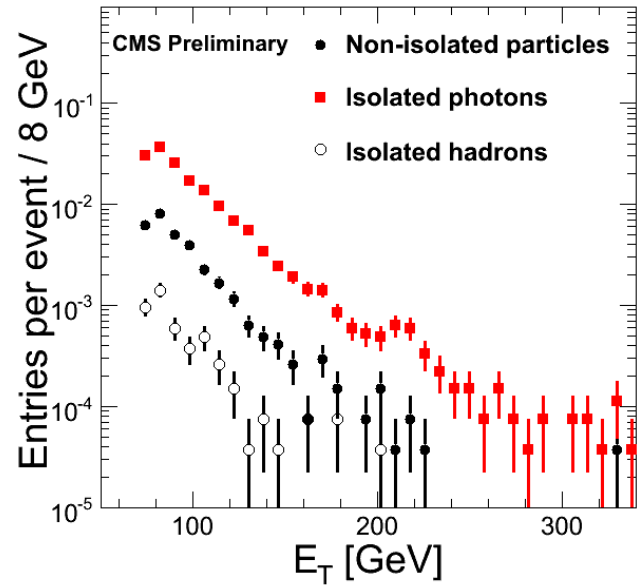}
\caption{Isolated photon spectrum as expected to be measured by CMS in one year of Pb-Pb collisions \cite{loizides1}.}
\label{figcmsgam}  
\end{figure}
CMS simulation studies are performed using PYQUEN~\cite{pyquen} and PYTHIA to generate the prompt photons with 
and without jet quenching respectively, and HYDJET~\cite{hydjet} to model the underlying heavy-ion event.
The working point of the photon detection is set to 60\% signal efficiency, 
leading to a background rejection of about 96.5\%.
The photon isolation and shape cuts improved S/B by a factor 15. 
A detailed description of the photon detection method can be found in~\cite{loizides1}.
The photon spectrum for an integrated luminosity of 0.5 nb$^{-1}$ is shown in  Fig.~\ref{figcmsgam}. 

\begin{figure}[htb]
\includegraphics[width=0.95\linewidth]{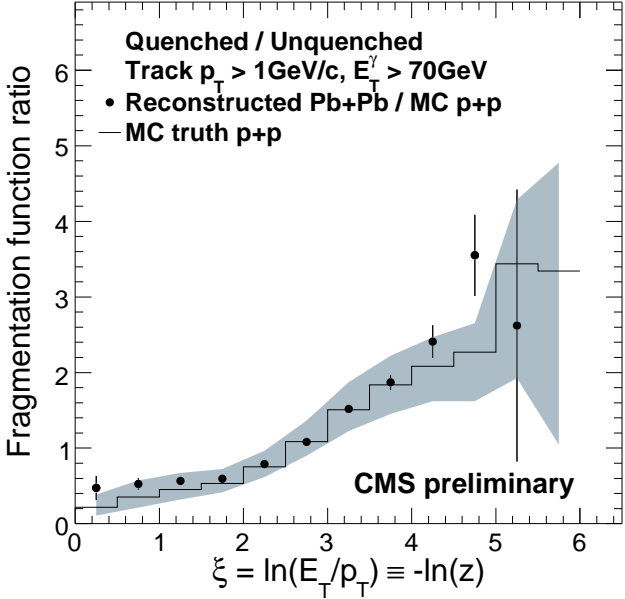}
\caption{Ratio of reconstructed (symbols) and MC truth (line) quenched fragmentation function over 
unquenched MC truth in CMS. The estimated systematic error is given by the shared band \cite{loizides1}.}
\label{figcmsrff}  
\end{figure}
Isolated photons are associated with the back-to-back jet in order to reconstruct the fragmentation function.
The ratio of the reconstructed fragmentation functions in quenched Pb-Pb and in pp is shown in Fig.~\ref{figcmsrff}.  
CMS obtains high significance for $\xi$ between 0.2 and 5 for E$_\gamma$ larger than 70 GeV (Fig.~\ref{figcmsrff}).

In addition to the $\gamma$-tagged jet measurement, jets can also be tagged with a virtual
photon or a Z$^0$. This method is free from the bias of the isolation cut that is applied in
$\gamma$-tagged jets. A detailed study for the CMS detector is presented in \cite{cms2,kunde}.

\section{Low mass dileptons}

The measurement of direct photons at low p$_T$ is usually performed on a statistical basis, namely
subtracting the decay photons from all detected photons \cite{pere1}. The difficulty of this me\-thod 
comes from the large background due to the decay photons. 
To overcome the background problem it has been proposed to consider instead the emission of virtual 
photons (lepton pairs). This method was used at CERN ISR to set a limit on direct photon 
production \cite{isr}. The invariant mass distribution of Dalitz decay pairs as well as of virtual photons 
is given by the Knoll-Wada formula \cite{knoll}. 

Beyond the mass of the $\pi^0$, only the $\eta$ mesons contributes. As most of the $\gamma$'s come from $\pi^0$ decays, 
the signal to background ratio for direct photon signal improves considerably for m$_{ee}>m_{\pi^0}$. 
This method also benefits from the excellent mass resolution at low
invariant mass and from the low conversion probability in the case of ALICE ($\sim$7\%). 
For ATLAS and for CMS, where the conversion probability is higher ($\sim$20\%, $\sim$30\%), 
it would be more difficult to apply this method.


This method has also been used recently by PHENIX to measure direct photons in pp and Au+Au \cite{phenix1}. 
While the yield is consistent with NLO pQCD calculations in pp collisions, 
in AuAu collisions the data are larger than calculations for p$_T<$3.5 GeV/c.

\begin{figure*}[hbt]
\begin{center}
\resizebox{0.85\textwidth}{!}{
  \includegraphics{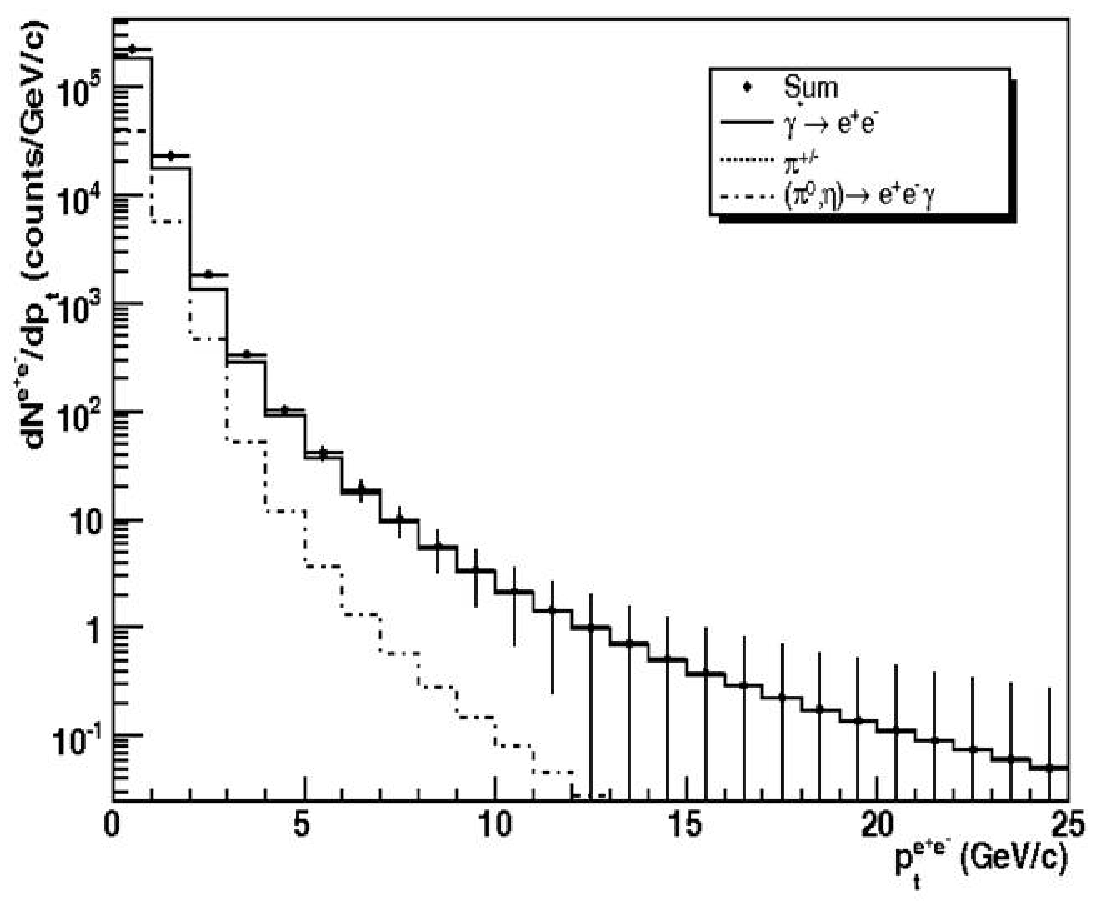}
  \includegraphics{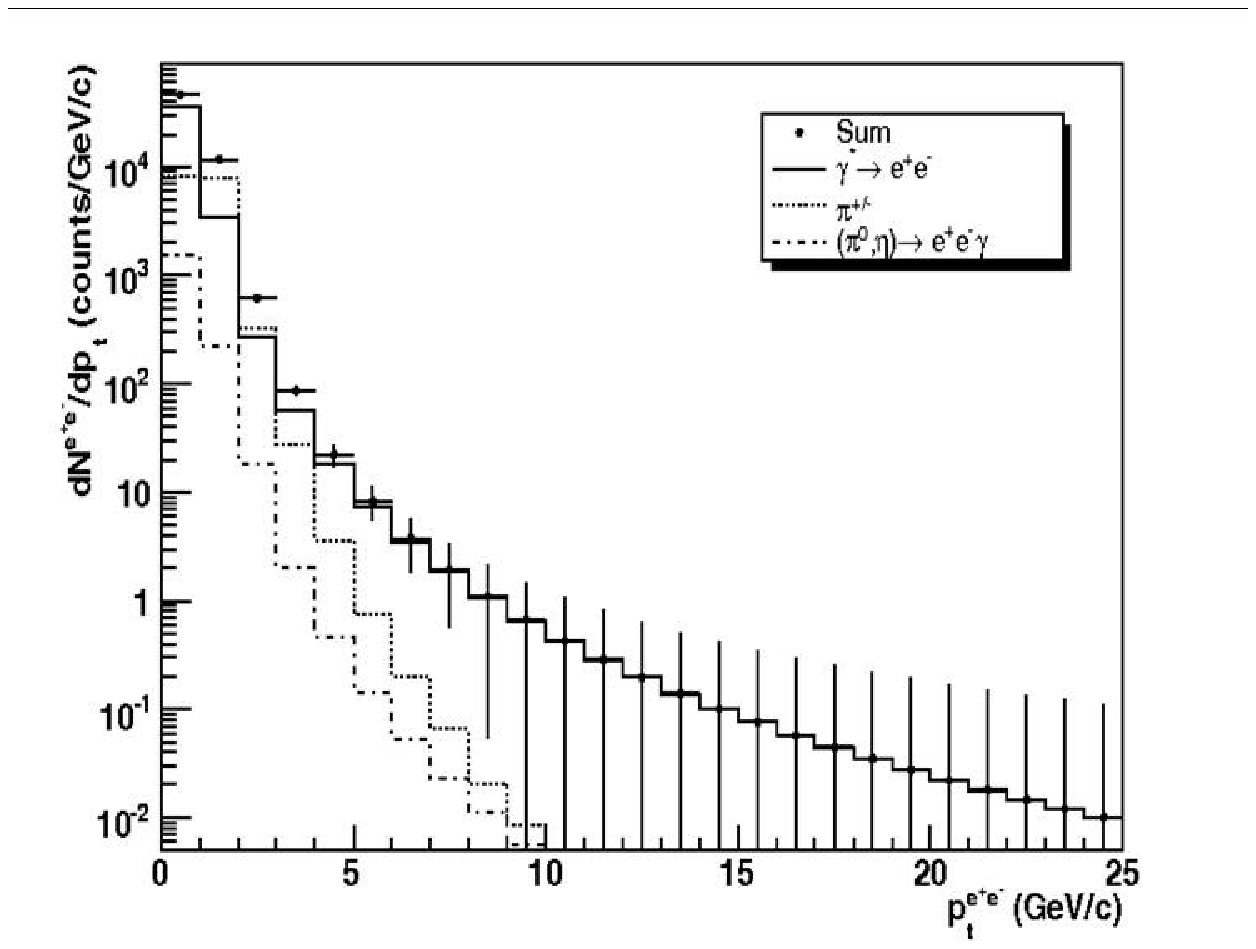}
}
\end{center}
\caption{Virtual photon spectrum reconstructed from the measurement of electron pairs with mass on the
range 0.2 to 0.6 GeV/c$^2$ in pp at $\sqrt{s}$ =5.5 TeV (left) and in central Pb-Pb collisions (right) 
at $\sqrt{s_{NN}}$=5.5 TeV.}
\label{figdilep}       
\end{figure*}

Theoretically, the rate of production of a dilepton pair in a finite mass range is given by \cite{aurenche}
\begin{equation}
{{d\sigma^{e^+e^-}} \over {dp_Tdy}} = C_{e^+e^-} \alpha  {{d\sigma^{\gamma}} \over {dp_Tdy}}
\label{eqee}
\end{equation}
where  C$_{e^+e^-}\sim 0.3$ for 0.2~GeV/c$^2<$M$_{e^+e^-}<$0.6~GeV/c$^2$ valid in the range 2 GeV/c$<$p$_T<$100 GeV/c.
The expected yield of virtual photons calculated taking into account Eq~\ref{eqee} and the pQCD rate for a LHC 
running year is shown in Fig.~\ref{figdilep}. Thermal photons are supposed to be the dominating contribution below 
p$_T$$\sim$5~GeV/c \cite{phot2,phot3,phot4}, therefore the dilepton yield is probably underestimated. 
Points for p$_T$ below 2 GeV/c should be taken with some caution, as this method assumes that the mass
of the dilepton pair is negligible compared to the p$_T$ and the validity of Eq.~\ref{eqee}.
The electron pairs are detected and identified by the ALICE central tracking system (TPC and TRD).

\section{Conclusions}

The LHC machine and the experiments are finally a reality. 
We have shown that the three experiments are equipped with very good photon and dilepton detectors 
that will allow the measurement of electromagnetic probes at LHC energies from low to very high p$_T$.
The low mass dilepton measurement will extend the direct photon spectrum to lower p$_T$.
The jet fragmentation function and its modification due to the medium effects in heavy-ion collisions
will be study using $\gamma$ tagged jets. 
The ALICE, ATLAS and CMS experiments will provide independent and complementary results,
that will be essential to understand this new system.

\begin{acknowledgement}
{\it Acknowledgements}. I would like to express my gratitude to the organizers of the 3rd 
International Conference on Hard and 
Electromagnetic Probes in High Energy Nuclear Collisions and the ALICE management for the invitation to
speak in A Toxa. I would like also to thank the ALICE, the ATLAS and the CMS 
Collaborations for providing the material for this talk and for useful discussions.
The author wishes to acknowledge the financial support received from GSI and from the German BMBF.
\end{acknowledgement}


\begin{thebibliography}{}

\bibitem{karsch1} F. Karsch, hep-lat/0106019.
\bibitem{lat2} M. Cheng {\it et al.}, Phys. Rev. D 77, 014511 (2008) 
\bibitem{fabjan1} C. Fabjan {\it et al.}, proceedings of Quark Matter Conference 2008
\bibitem{phot1} S. Turbide {\it et al}, Phys. Rev. C 72, 014906 (2005) 
\bibitem{phot2} R.J. Fries {\it et al}, Phys. Rev. Lett. 90, 132301 (2003) 
\bibitem{phot3} S. Turbide {\it et al}, Phys. Rev. C 69, 014903 (2004) 
\bibitem{phot4} F. Arleo {\it et al}, nucl-th/0707.2357v2.
\bibitem{armesto} N. Armesto {\it et al}, J. Phys. G: Nucl. Part. Phys. 35, 054001 (2008) 
\bibitem{photyellow} F. Arleo  {\it et al}, LHC CERN Yellow Report 2004-009. hep-ph/0311131
\bibitem{turbide1} S. Turbide {\it el al}, Phys. Rev. Lett. 96, 032303 (2006) 
\bibitem{liu1} W. Liu, R.J. Fries, nucl-th/0801.0453.
\bibitem{klein1} C. Klein-Boesing {\it et al} (ALICE Collaboration), these proceedings.
\bibitem{loizides1} C. Loizides {\it et al}, (CMS Collaboration), arXiv 0804.3679v1, nucl-ex, proceedings of Quark Matter Conference 2008
\bibitem{grau1} N. Grau {\it et al}, (ATLAS Collaboration), arXiv:0805.4656v1, nucl-ex
\bibitem{ali1} ALICE Collaboration {\it et al}, J. Phys. G: Nucl. Part. Phys. 30, 1517 (2004) 
\bibitem{ali2} ALICE Collaboration {\it et al}, J. Phys. G: Nucl. Part. Phys. 32, 1295 (2006) 
\bibitem{atlas1} ATLAS Collaboration, ATLAS Technical Design Report, vol 1 CERN/LHCC 1999-14, vol 2  CERN/LHCC 1999-15
\bibitem{cms1} CMS Collaboration {\it et al}, The CMS Physics Technical Design Report, Volume 1, CERN/LHCC 2006-001
\bibitem{atlas2} ATLAS Collaboration, Letter~of~Intent, Report CERN/LHCC/2004-009 
\bibitem{cms2} D.G.~d'Enterria {\it et al} (CMS Collaboration), J. Phys. G: Nucl. Part. Phys. 34, 2307 (2007)
\bibitem{trd} ALICE Collaboration {\it al}, TRD Technical Design Report, CERN/LHCC 2001-021
\bibitem{pythia} T. Sjostrand {\it et al}, Comp. Phys. Comm. 135, 238 (2001) 
\bibitem{hijing} X.N. Wang , N. Gyulassy, Phys. Rev. D 44 (1991) 3501; M. Gyulassy, X.N. Wang, Comp. Phys. Comm. 83, 307 (1994) 
\bibitem{nimali1} G. Conesa {\it et al}, Nucl. Instr. and Meth. A 580, 1446 (2007) 
\bibitem{nimali2} G. Conesa {\it et al}, Nucl. Instr. and Meth. A 585, 28 (2008) 
\bibitem{yaxian1} Y. Mao {\it et al}, (ALICE Collaboration), proceedings of Quark Matter Conference 2008. ALICE-INT (2007) 021
\bibitem{cole1} B. A. Cole {\it et al}, (ATLAS Collaboration), these proceedings
\bibitem{pyquen} I. P. Lokhtin and A. M. Snigirev, Eur. Phys J. C 45, 211 (2006) [PYQUEN v1.2 is used] 
\bibitem{hydjet} I. P. Lokhtin and A. M. Snigirev, ArXiv:hep-ph/0312204 [HYDJET v1.2 is used] 
\bibitem{kunde} C. J. Kunde {\it et al}, these proceedings
\bibitem{pere1} D. Peressounko {\it et al}, (ALICE Collaboration), these proceedings
\bibitem{isr} J. H. Cobb {\it et al}, Phys. Lett. B 78, 519 (1978) 
\bibitem{knoll} N. M. Knoll and W. Wada, Phys. Rev. 98, 1355 (1955) 
\bibitem{phenix1} A. Adare {\it et al}, (PHENIX Collaboration) arXiv:0804.4168v1 (nucl-ex).
\bibitem{aurenche} P. Aurenche {\it et al}, Phys. Lett. B 209, 375 (1988) 

\end{thebibliography}
\end{document}